\documentstyle[12pt,epsfig]{article}
\begin{document}
%\input{psfig}
%\begin{titlepage}
\pagestyle{myheadings}
%\begin{flushright} 
%{BROWN-HET-} \\
%{September 1995}
%\end{flushright} 
\vspace*{0.8cm}
 \begin{center} 
{\large \bf Jacobi Elliptic Solutions of $\lambda\phi^4$ Theory} \\
{\large \bf in a Finite Domain}
\\ [10mm]
\end{center} 
\renewcommand{\thefootnote}{\alph{footnote}} 
\begin{center}
J. A. Espich\'an Carrillo$^{a}$ %\footnote{e-mail:espichan@ifi.unicamp.br}\\ 

{\it Instituto de F\'{\i}sica ``Gleb Wathagin'', University of Campinas
(UNICAMP) - 13.081-970 - Campinas (SP), Brazil.}
\end{center}
 
\begin{center}
A. Maia Jr.$^{b}$ %\footnote{e-mail:maia@ime.unicamp.br}\\

{\it Instituto de Matem\'atica, University of Campinas (UNICAMP) - 
13.081-970 - Campinas (SP), Brazil.}
\end{center}

\begin{center}
V. M. Mostepanenko$^{c}$ %\footnote{e-mail:mostep@fisica.ufpb.br}\\

{\it Friedmann Laboratory for Theoretical Physics (Russia) and Department of 
Physics, Federal University of Para\'{\i}ba (UFPB) - C.P. 5008; 58059-970 - 
Jo\~ao Pessoa (PB), Brazil.}
\end{center}

\begin{center}
\vspace{1.0cm} 
{\bf
Abstract}
 \end{center} 
  
\vspace*{0.8cm}

\noindent
The general static solutions of the scalar field equation for the potential 
$V(\phi)= -\frac{1}{2}M^2\phi^2 + \frac{\lambda}{4}\phi^4$ are determined for 
a finite domain in $(1+1)$ dimensional space-time. A
family of real solutions is described in terms of Jacobi Elliptic Functions.  
We show that the vacuum-vacuum boundary 
conditions can be reached by elliptic {\rm cn}-type solutions in a finite 
domain, such as of the Kink, for which they are imposed at infinity. We proved 
uniqueness for elliptic {\rm sn}-type solutions
satisfying Dirichlet boundary conditions in a finite interval (box) as well 
the existence of a minimal mass corresponding to these solutions in a box.

\noindent
We define expressions for the ``topological charge'', 
``total energy'' (or classical mass) and ``energy density'' for elliptic 
{\rm sn}-type solutions in a finite domain. For large length of the box
the conserved charge, classical mass and
energy density of the Kink are recovered. Also, we have shown that using 
periodic boundary conditions the results are the same as in the case of 
Dirichlet boundary conditions. In the case of anti-periodic boundary conditions
all elliptic {\rm sn}-type solutions are allowed.

%\noindent
%We show that the Kink (Anti-Kink) is the unique solution which satisfies
%the
%minimal requirements of finiteness and regularity (class $C^{2}(\Re))$.

\vspace*{0.6cm}
\noindent
$^{a}$E-mail: espichan@ifi.unicamp.br\\
$^{b}$E-mail: maia@ime.unicamp.br\\
$^{c}$E-mail: mostep@fisica.ufpb.br

\baselineskip 0.65cm

\newpage

\noindent
{\Large \bf 1. General Solutions}

\vspace*{0.5cm}

In this paper we obtain static solutions of the 
classical equation of motion of a scalar field $\phi$, with potential 
$V(\phi)= -\frac{1}{2}M^2\phi^2 + \frac{\lambda}{4}\phi^4$ in a finite
domain. Static solutions with vacuum-vacuum boundary conditions at 
$x= \pm\infty$ were firstly obtained by Dashen et all (DHN)$^{1}$.
However, our case may be important since, as far as we know, no general 
solution satisfying the same boundary conditions
for the above potential was found in the literature in a finite 
domain. Also this has an evident connection with the Casimir effect 
which attracted much attention for its applications in several topics
of physics $^{2}$.

Let us consider the Lagrangian density of a scalar field theory in $(1+1)$
dimensions given by
\begin{eqnarray*}
L(\phi,\partial_{\mu}\phi)=\frac{1}{2}\dot{\phi^2} -
\frac{1}{2}(\partial_{x}\phi)^{2}-V(\phi),
\end{eqnarray*}
where $V(\phi) = -\frac{1}{2}M^2\phi^2 + \frac{\lambda}{4}\phi^4$,
$M$ is the mass of the field and the dot means derivative with respect to $t$.

The extrema of the potential $V(\phi)$ are reached for
\begin{eqnarray*}
\phi_{e} = 0, \ \ \ \ \ \ \phi_{m}=\pm\frac{M}{\sqrt{\lambda}}\ \mbox{(vacua)}.
\end{eqnarray*}

We are interested in the static case $\phi=\phi(x)$. So the classical equation 
of motion is
\begin{eqnarray}
\partial_{xx}\phi-\frac{\partial V}{\partial\phi}=0.
\label{sn1}
\end{eqnarray}

It is well known that the solutions for Eq. (\ref{sn1}), 
satisfying vacuum to vacuum boundary conditions at $\pm\infty$, are given by
\begin{eqnarray}
\phi_0(x) = \pm\frac{M}{\sqrt{\lambda}}tanh\left(\frac{M}{\sqrt{2}}(x - x_0)
\right).
\label{sn2}
\end{eqnarray}

These solutions are called ``Kink''($+$ sign) and ``Anti-Kink''($-$ sign). 
The solution $\phi_0$ approaches to
two different values $\pm\frac{M}{\sqrt{\lambda}}$ as $x\rightarrow\pm\infty$, 
which correspond to a
degenerate vacuum configuration.

We now turn to the calculation of the static solutions of the classical equation 
of motion (\ref{sn1}) with boundary conditions in a finite domain which, in 
this paper, is a bounded interval. In order to have well behaved solutions
we impose our solutions belonging to class $C^{2}(\Omega)$, where 
$\Omega\subset\Re$ is an interval.

A first integral of Eq. (\ref{sn1}), 
after a change of variable, is given by
\begin{eqnarray}
x -x_{0} = \pm \frac{1}{M}\,\int\frac{dz}{\sqrt{z^4 - z^2 + \frac{c}{2}}},
\label{sn3}
\end{eqnarray}
where $z = \frac{\sqrt{\lambda}}{\sqrt{2}M}\phi$ is the dimensionless variable
\ and \ $x_{0}$, $c$ are constants of integration.

Since we are
interested to find real solutions of Eq. (\ref{sn3}), we
must evaluate the integral of right-hand side for
$(z^4 - z^2 + \frac{c}{2}) >0$.

For any $c$, the stationary points of the function 
$f(z)=\frac{1}{\sqrt{z^4 - z^2 + \frac{c}{2}}}$ are given by the values:
\begin{eqnarray*}
z=0, \ \ \pm\frac{1}{\sqrt{2}}.
\end{eqnarray*}

On the other hand, note that for $c=0$ and $c=\frac{1}{2}$, the integral in Eq. 
(\ref{sn3}) is not defined at these points. So, we 
discuss the following cases:
\begin{eqnarray*}
1) \ c\leq 0, \ \ \ \ \ \ 2) \ 0<c<\frac{1}{2} \ \ \ \ \ \mbox{and} \ \ \ \ \ 
3) \ c\geq \frac{1}{2}.
\end{eqnarray*}

It is important to emphasize that we are interested to find static solutions 
$\phi(x)$  that satisfy the vacuum-vacuum boundary conditions, like the 
solution (\ref{sn2}), but now in a finite domain and to study some of their
properties.

\vspace*{0.5cm}

1) {\bf CASE $c\leq 0$}

\vspace*{0.2cm}

In this case $(z^4 - z^2 + \frac{c}{2}) >0$ is satisfied for
$|z|>\sqrt{\frac{1+\sqrt{1-2c}}{2}}$.

The integral in Eq. (\ref{sn3}) can be performed to 
give $^{3}$
\begin{eqnarray}
\int\frac{dz}{\sqrt{z^4 - z^2 + \frac{c}{2}}}=\frac{1}{\sqrt[4]{1-2c}}\,
F(\theta,m),
\label{sn4}
\end{eqnarray}
where
\begin{eqnarray*}
F(\theta,m)=\int\limits_{0}^{\theta}\frac{dt}{\sqrt{1-msin^{2}t}}
\end{eqnarray*}
is an elliptic integral of the first kind, and
\begin{eqnarray*}
\theta = arccos(\frac{\sqrt{1+\sqrt{1-2c}}}{\sqrt{2}z}), \ \ \
\ \ m = \frac{-1+ \sqrt{1-2c}}{2\sqrt{1-2c}}.
\end{eqnarray*}

So, the Eq. (\ref{sn3}) is equivalent to
\begin{eqnarray}
F\left( arccos(\frac{\sqrt{1+\sqrt{1-2c}}}{\sqrt{2}z}), 
\frac{-1+ \sqrt{1-2c}}{2\sqrt{1-2c}}\right) =\pm M\sqrt[4]{1-2c}\,(x-x_0).
\label{sn5}
\end{eqnarray}

Solving (\ref{sn5}) for $z$ we get (we substitute $z$ by 
$\frac{\sqrt{\lambda}}{\sqrt{2}M}\phi$)
\begin{eqnarray}
\phi_{c}(x)=\frac{M\sqrt{1+\sqrt{1-2c}}}{\sqrt{\lambda}}\,
\frac{1}{{\rm cn}\left(\pm \sqrt[4]{1-2c}\,M\,(x-x_0),\frac{-1+\sqrt{1-2c}}
{2\sqrt{1-2c}}\right)},
\label{sn6}
\end{eqnarray}
where ${\rm cn}(u,m)$ is a Jacobi Elliptic Function
$^{4,}$\footnote[1]{Jacobi Elliptic Functions can be used to describe 
solutions of other non-linear equations such as KdV equation. $^{5}$.}
The solution (\ref{sn6}) is unbounded because ${\rm cn}$ has zeros.
Observe that the functions ${\rm cn}^{-1}(u,m)$ 
are larger than or equal to $1$ and since 
$\sqrt{1+\sqrt{1-2c}}\geq \sqrt{2}$, then
the value of $\phi_{c}(x)$ is always larger than 
$\frac{M}{\sqrt{\lambda}}$. So the solutions given by (\ref{sn6}) can not 
satisfy vacuum-vacuum boundary conditions.

\vspace*{0.5cm}

2) {\bf CASE $0<c<\frac{1}{2}$}

\vspace*{0.2cm}

Here $(z^4 - z^2 + \frac{c}{2}) >0$ is satisfied for
the following cases
\begin{eqnarray*}
|z|\geq\sqrt{\frac{1+\sqrt{1-2c}}{2}} \ \ \ \ \ \ \mbox{and} \ \ \ \ \ \ 
|z|\leq\sqrt{\frac{1-\sqrt{1-2c}}{2}}.
\end{eqnarray*}

\vspace*{0.5cm}

a) $|z|\geq\sqrt{\frac{1+\sqrt{1-2c}}{2}}$

\vspace*{0.2cm}

The integral in Eq. (\ref{sn3}) is given by $^{3}$
\begin{eqnarray}
\int\frac{dz}{\sqrt{z^4 - z^2 + \frac{c}{2}}}=-\frac{\sqrt{2}}
{\sqrt{1+\sqrt{1-2c}}}\,F\left( arcsin(\frac{\sqrt{1+\sqrt{1-2c}}}{\sqrt{2}z}),
\frac{1}{-1+\frac{1+\sqrt{1-2c}}{c}}\right).
\label{sn7}
\end{eqnarray}

Thus, Eq. (\ref{sn3}) can be rewritten as
\begin{eqnarray}
F\left( arcsin(\frac{\sqrt{1+\sqrt{1-2c}}}{\sqrt{2}z}),\frac{1}{-1+
\frac{1+\sqrt{1-2c}}{c}}\right)=\mp\,M\sqrt{\frac{1+\sqrt{1-2c}}{2}}\,(x-x_0).
\label{sn8}
\end{eqnarray}

As in the previous case the solution of the Eq. (\ref{sn8}) is
given by 
\begin{eqnarray}
\phi_{c}(x)=\frac{M\sqrt{1+\sqrt{1-2c}}}{\sqrt{\lambda}}\,
\frac{1}{{\rm sn}\left(\pm \sqrt{\frac{1+\sqrt{1-2c}}{2}}\,M\,(x-x_0),\frac{1}
{-1+\frac{1+\sqrt{1-2c}}{c}}\right)},
\label{sn9}
\end{eqnarray}
where ${\rm sn}(u,m)$ is a Jacobi Elliptic Function $^{4}$.
As in the previous case, this solution is unbounded since ${\rm sn}$ 
also has zeros.

It is easy to see that the amplitude of $\phi_{c}(x) > \frac{M}
{\sqrt{\lambda}}$. Therefore the solutions given by (\ref{sn9}) also can not 
satisfy vacuum-vacuum boundary condition. 
Observe that, for the limit $c\rightarrow\frac{1}{2}$, we obtain from (\ref{sn9}) 
the solution
\begin{eqnarray*}
\phi(x) = \pm\frac{M}{\sqrt{\lambda}}coth\left(\frac{M}{\sqrt{2}}(x - x_0)
\right).
\end{eqnarray*}
Although this solution reaches the vacuum-vacuum boundary conditions (at 
$\pm\infty$)
it has a discontinuity at $ x = x_{0}$ and therefore must be discarded.

\newpage

%\vspace*{0.7cm}

b) $|z|\leq\sqrt{\frac{1-\sqrt{1-2c}}{2}}$

In this case we have $^{3}$
\begin{eqnarray}
\int\frac{dz}{\sqrt{z^4 - z^2 + \frac{c}{2}}}=\frac{\sqrt{2}}
{\sqrt{1+\sqrt{1-2c}}}\,F\left( arcsin(\frac{\sqrt{2}z}{\sqrt{1-\sqrt{1-2c}}}),
\frac{1}{-1+\frac{1+\sqrt{1-2c}}{c}}\right).
\label{sn10}
\end{eqnarray}

From (\ref{sn10}) and (\ref{sn3}) we get
\begin{eqnarray}
F\left( arcsin(\frac{\sqrt{2}z}{\sqrt{1-\sqrt{1-2c}}}),
\frac{1}{-1+\frac{1+\sqrt{1-2c}}{c}}\right)=\pm\,\frac{M\sqrt{1+\sqrt{1-2c}}}
{\sqrt{2}}(x-x_0).
\label{sn11}
\end{eqnarray}

The solution of the Eq. (\ref{sn11}) is
given by 
\begin{eqnarray}
\phi_{c}(x) = \pm\frac{M}{\sqrt{\lambda}}\frac{\sqrt{2c}}{\sqrt{1+\sqrt{1-2c}}}\,
{\rm sn}\left(\frac{M\sqrt{1+\sqrt{1-2c}}}{\sqrt{2}}(x-x_0),\frac{1}
{-1+\frac{1+\sqrt{1-2c}}{c}}\right).
\label{sn12}
\end{eqnarray}

This is a family of {\rm sn}-type elliptic functions
parametrized by the constant of integration $c$.
Observe that, although we are not considering it by now, the Kink
solution (\ref{sn2}) is re-obtained as a limiting case 
$c\rightarrow\frac{1}{2}$.

\begin{figure}[h]
\centerline{\psfig{figure=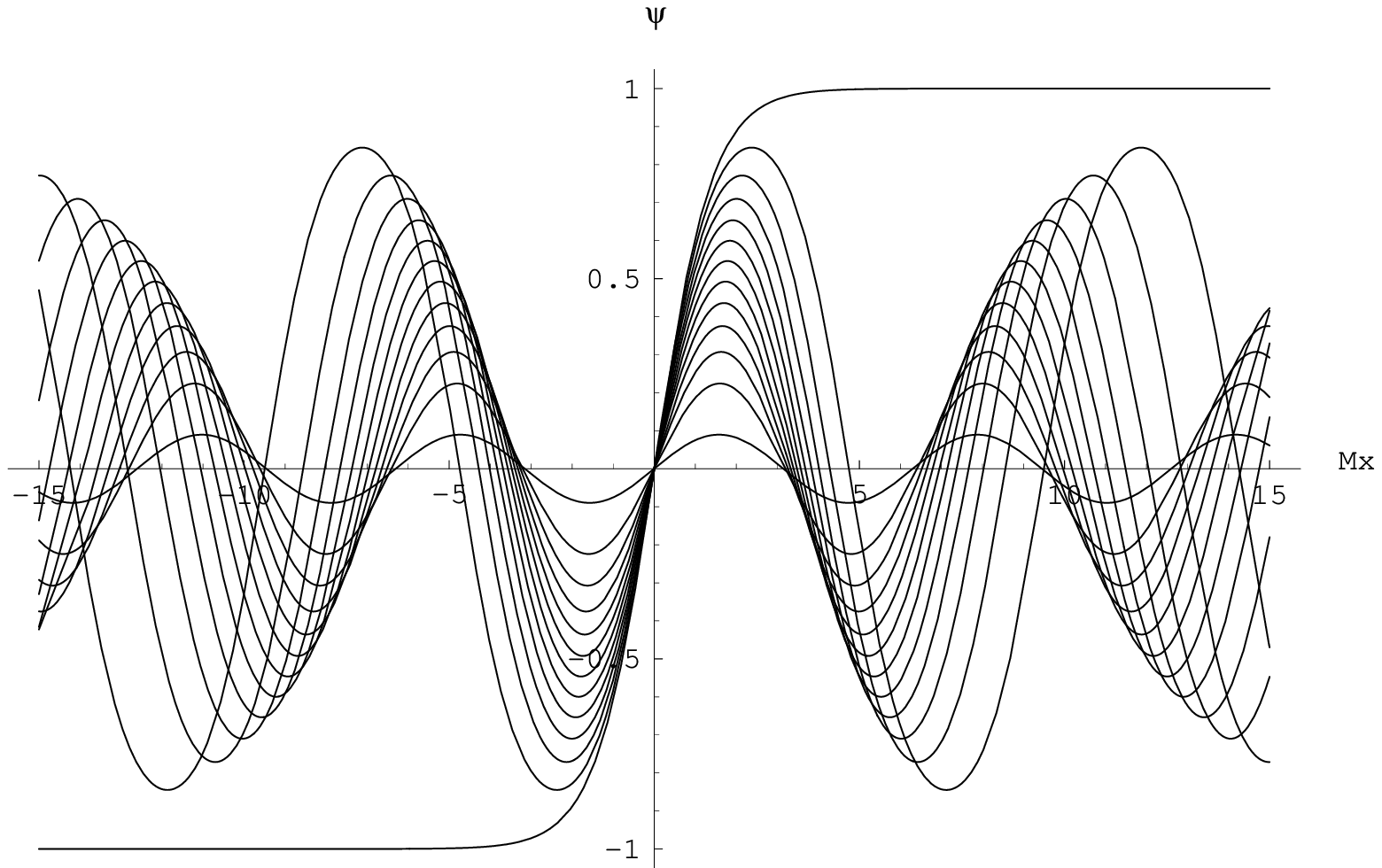,height=3.3in}}
%\caption{}
\end{figure}
\begin{center}
\noindent
\parbox{13cm}{Fig. 1. Family of {\rm sn} elliptic functions for c = 0.008, 0.049, 0.09, ...,
0.418, 0.459, 0.5. The solution for c = 0.5 is the DHN's Kink. Here and below 
we have defined $\psi = \frac{M}{\sqrt{\lambda}}\phi$.}
\end{center}

\vspace*{0.3cm}

Next, we show that the  vacuum-vacuum boundary condition can not be
satisfied by any
solution (\ref{sn12}), unless $c = \frac{1}{2}$. For the Jacobi Elliptic 
Functions we have $|{\rm sn}(u,m)| < 1 $  and since 
$\sqrt{1-\sqrt{1-2c}} < 1$, also $|\phi_{c}(x)| <\frac{M}{\sqrt{\lambda}}$. 
So the general solution (\ref{sn12}) can not reach the boundary condition 
vacuum-vacuum. 
Therefore we proved that there are no solutions of Kink-type for 
$0<c<\frac{1}{2}$. Only the kink solution ($c=\frac{1}{2}$)
satisfies vacuum to vacuum boundary 
condition at $\pm \infty$ among the solutions with $c\leq\frac{1}{2}$.

\vspace*{0.8cm}

3) {\bf CASE $c\geq\frac{1}{2}$}

In this case $(z^4 - z^2 + \frac{c}{2}) >0$ is satisfied for any $z$.
So, we have $^{3}$
\begin{eqnarray}
\int\frac{dz}{\sqrt{z^4 - z^2 + \frac{c}{2}}}=\frac{1}{2}\sqrt[4]
{\frac{2}{c}}\,F\left( 2arctan(\sqrt[4]{\frac{2}{c}}z),
\frac{1}{2}(1+\frac{1}{\sqrt{2c}})\right).
\label{sn13}
\end{eqnarray}

Substituting (\ref{sn13}) in (\ref{sn3}) we obtain
\begin{eqnarray}
F\left( 2arctan(\sqrt[4]{\frac{2}{c}}z),
\frac{1}{2}(1+\frac{1}{\sqrt{2c}})\right)=\pm\,2M\sqrt[4]{\frac{c}{2}}(x-x_0).
\label{sn14}
\end{eqnarray}

The solution of the Eq. (\ref{sn14}) is
\begin{eqnarray}
|\phi_{c}(x)| = \frac{M\sqrt[4]{2c}}{\sqrt{\lambda}}\,\left|
\frac{{\rm sn}
\left(\sqrt[4]{\frac{c}{2}}\,M(x-x_0),\frac{1}{2}(1+\frac{1}{\sqrt{2c}})\right)
{\rm dn}\left(\sqrt[4]{\frac{c}{2}}\,M(x-x_0),\frac{1}{2}(1+\frac{1}{\sqrt{2c}})
\right)}{{\rm cn}\left(\sqrt[4]{\frac{c}{2}}\,M(x-x_0),\frac{1}{2}(1+\frac{1}
{\sqrt{2c}})\right)}\right|.
\label{sn15}
\end{eqnarray}

Due to the identity
\begin{eqnarray*}
\frac{{\rm sn}x\,{\rm dn}x}{{\rm cn}x}=\left(\frac{1-{\rm cn}2x}{1+
{\rm cn}2x}\right)^{\frac{1}{2}}
\end{eqnarray*}
this solution may be called {\rm cn}-type solution.

Since above solution is periodic, it is easy to see that $\phi_{c}(x)$  
has an infinite number of branches for each value of $c$.
In Fig(s) 2 and 3 we show, respectively, the
cases $c=\frac{1}{2}$ and $c=1$. Notice that in the case $c=1$ we show
just four possibilities for the solutions (\ref{sn15}) in the real line. In
general, for each period of the function we get four different cases. When we 
consider all possibilities for each period an infinite number of branches is 
obtained. Nevertheless, for our purposes only the behaviour of the function 
inside one period is significant. So we restrict our considerations below to 
the interval $ (- 2 K(m'), 2 K(m'))$,
where
\begin{eqnarray*}
K(m') = \int\limits_{0}^{\frac{\pi}{2}}
\frac{dt}{\sqrt{1-m'\,sin^{2}t}} 
\end{eqnarray*}
is a Complete Elliptic Integral of the first kind and 
\begin{eqnarray*}
m' = \frac{1}{2}(1+\frac{1}{\sqrt{2c}}).
\end{eqnarray*}

For $c=\frac{1}{2}$ the Kink and Anti-Kink solutions are recovered (see Fig. 2).
However, for the solutions given by fig(s). 2.1 and 2.2 the situation is 
different, since they satisfy the condition vacuum-vacuum, but for the same 
vacuum. 
Also these solutions have a discontinuity for the first derivative at 
$x=0$. This discontinuity in the solutions of this type will appear for any 
lenght $L$. So, these solutions does not belong to the class $C^{2}(\Re)$ we 
are considering for solving Eq. (\ref{sn1}). Therefore we discard them as 
acceptable solutions.

\begin{figure}[ht]
\centerline{\psfig{figure=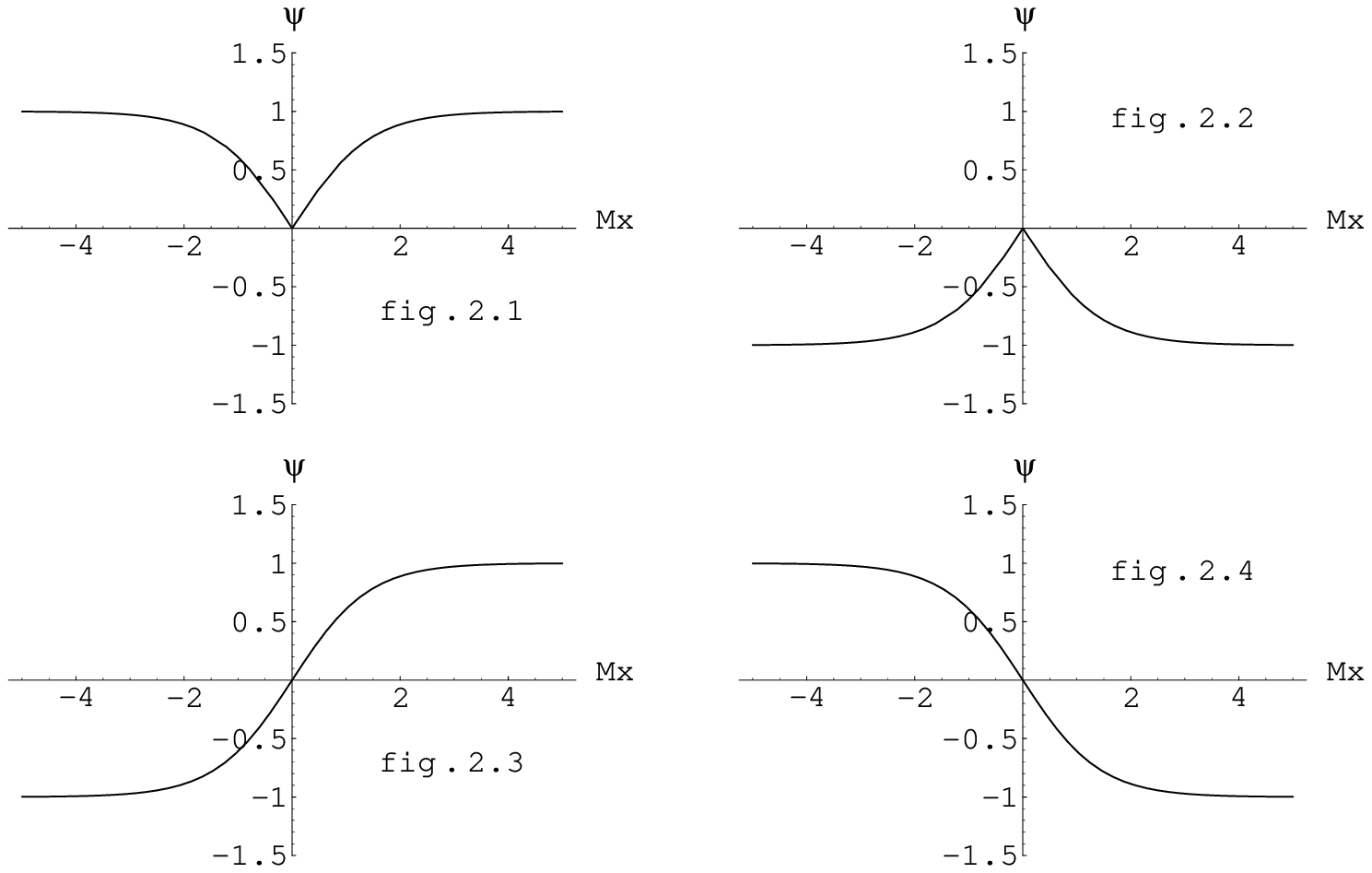,height=3.3in}}
%\caption{}
\end{figure}
\begin{center}
\noindent
\parbox{13cm}{Fig. 2. Solutions given by Eq. (\ref{sn15}) for $c = \frac{1}{2}$. The  
Kink and AntiKink solutions are recovered in this case. For the first two 
functions $\phi'$ is not continuous at $x=0$, so they are not admissible.}
\end{center}

\vspace*{0.5cm}

From Fig(s) 3.3 and 3.4 an interesting case can be studied. If we impose 
vacuum-vacuum boundary conditions 
$\phi(\frac{L}{2})= \phi(-\frac{L}{2}) = \frac{M}{\sqrt{\lambda}}$
for a fixed length $L$ we obtain from (\ref{sn15}) (for $c=1$, for example), 
the following relation
\begin{eqnarray*}
\frac{{\rm sn}
\left(\frac{ML}{2\sqrt[4]{2}},0.85\right)
{\rm dn}\left(\frac{ML}{2\sqrt[4]{2}},0.85\right)}
{{\rm cn}\left(\frac{ML}{2\sqrt[4]{2}},0.85\right)}\ = \ 0.84.
\end{eqnarray*} 

This result can be generalized for $c$ arbitrary, that is,
\begin{eqnarray}
\frac{{\rm sn}
\left(\sqrt[4]{\frac{c}{2}}\frac{ML}{2},\frac{1}{2}(1+\frac{1}{\sqrt{2c}})\right)
{\rm dn}\left(\sqrt[4]{\frac{c}{2}}\frac{ML}{2},\frac{1}{2}(1+\frac{1}{\sqrt{2c}})
\right)}{{\rm cn}\left(\sqrt[4]{\frac{c}{2}}\frac{ML}{2},\frac{1}{2}(1+\frac{1}
{\sqrt{2c}})\right)}\ = \ \frac{1}{\sqrt[4]{2c}}.
\label{sn16}
\end{eqnarray}

Therefore, this case (Fig(s) 3.3 and 3.4) shows that the vacuum-vacuum boundary 
condition can be satisfied on a finite domain in the spirit of
Casimir Effect, but here in the classical level. 
For the case in the Fig(s) 3.1 and 3.2 the condition vacuum-vacuum is 
satisfied for the same vacuum. 
Moreover, as in the previous case ($c = \frac{1}{2}$) these solutions have
a discontinuity for the first derivative at $x=0$. So, also here we discard
them as acceptable solutions.

\begin{figure}[ht]
\centerline{\psfig{figure=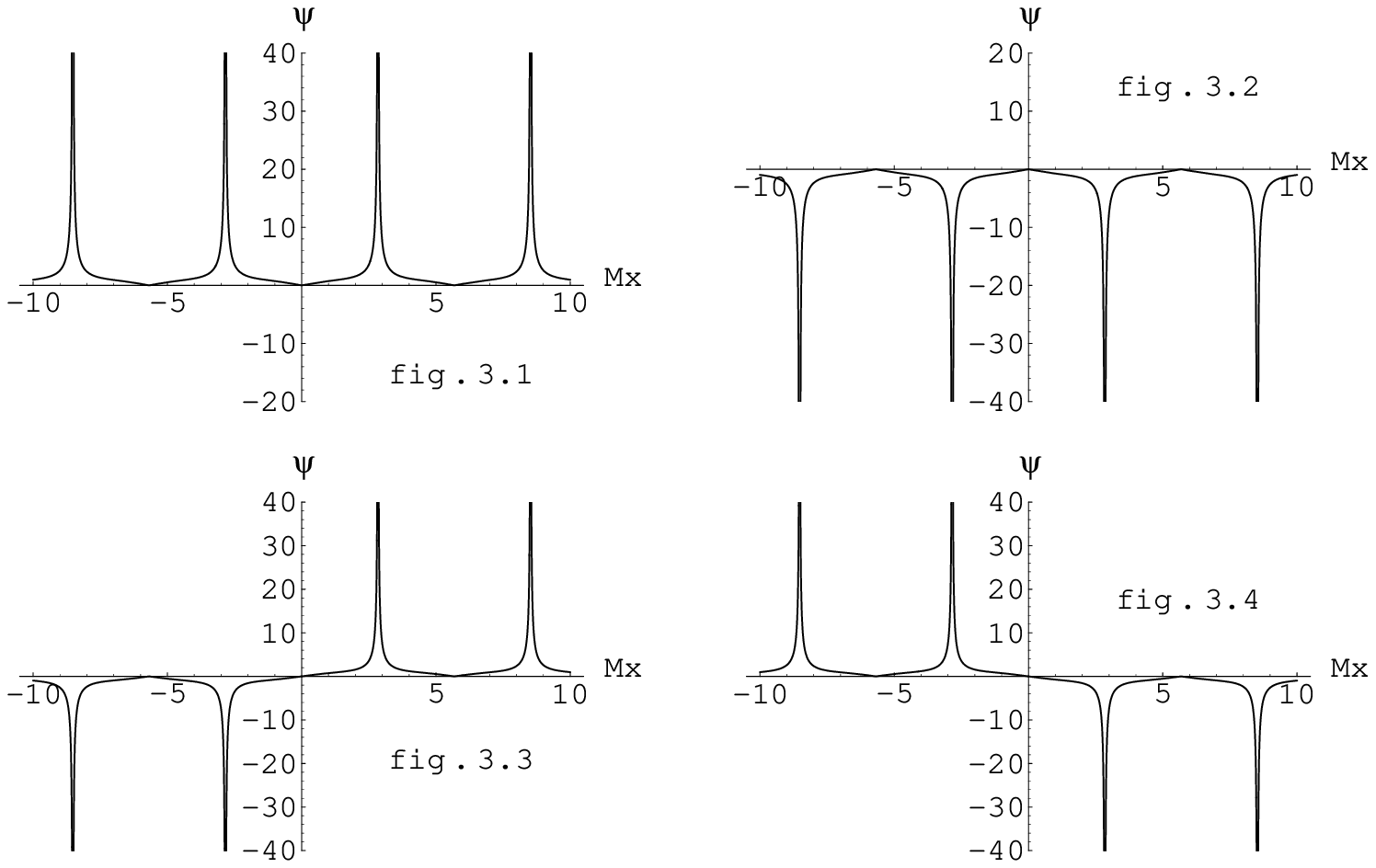,height=3.3in}}
%\caption{}
\end{figure}
\begin{center}
\noindent
\parbox{13cm}{Fig. 3. Solutions given by Eq. (\ref{sn15}) for $c = 1$. Here
we took $x_0 =0$. For a given length $L$ there exists just one
solution satisfying vacuum-vacuum condition at $\pm\frac{L}{2}$.}
\end{center}

\vspace*{1cm}

\noindent
{\Large \bf 2. Properties of {\rm sn}-Type Solutions in an One-Dimensional Box}

\vspace*{0.7cm}

Now, we study the bounded {\rm sn}-type
solutions of class $C^{2}(\Omega)$  
(Eq. (\ref{sn12})) that satisfy the Dirichlet boundary conditions.

First of all, it is clear that
translation invariance is not valid any more due to Dirichlet
boundary conditions on a finite domain, that is, a translated {\rm sn} in 
general does not satisfy the same boundary conditions (in our case Dirichlet's)
in the same interval. Only the Kink $(c=\frac{1}{2})$ has this property in
the infinite interval $(-\infty ,+\infty )$.

Now we are going to show that in a given interval 
$[-\frac{L}{2},+\frac{L}{2}]$
there exists only one {\rm sn} satisfying Dirichlet boundary conditions at
$x = \pm\frac{L}{2}$. The proof is as follows.
We note that, in order a solution of Eq. (\ref{sn12})
satisfies Dirichlet boundary conditions, that is $\phi_{c}(\pm \frac{L}{2})=0$, 
we must have
\begin{eqnarray}
{\rm sn}\left(\frac{ML}{2\sqrt{2}}\sqrt{1+\sqrt{
1-2c}},\frac{1}{-1+\frac{1+\sqrt{1-2c}}{c}}\right) = 0.
\label{sn17}
\end{eqnarray}
(we take, without loss of generality, $x_0=0$).

We have from the theory of Jacobi Elliptic Functions
\begin{eqnarray}
\frac{ML}{2\sqrt{2}}\sqrt{1+\sqrt{1-2c}}=2\,K_{n}(m),
\label{sn18}
\end{eqnarray}
where 
\begin{eqnarray*}
K_{n}(m)=\frac{K(m)}{n}, \ \ n=1,2,3\dots ,\ \ \ \ \ \ \ \mbox{and} \ \ \ \ \ \ 
m= \frac{1}{-1+\frac{1+\sqrt{1-2c}}{c}}.  
\end{eqnarray*}

Here $4\,K(m)$ 
is a period of the Jacobi Elliptic Functions ${\rm sn}(u,m)$ $^{4}$. 

On the other hand, since our solutions are Jacobi Elliptic Functions we must 
have  $K_{n}(m)\geq\frac{\pi}{2}$. This is true only for $n=1$. So, there is no 
solution of the $sn$-type  with 
semi-period less than $2K(m)$. 

Also we observe from (\ref{sn12}) that 
\begin{eqnarray*}
\lim_{c\rightarrow 0}\,\mid\phi_{c}(x)\mid =0 \ \ \ \ \ \ \forall x\in\Re.
\end{eqnarray*}

For fixed $x\in\Re$, but large, the convergence of the limit above is very slow.
 
Now, with $n=1$, we obtain from (\ref{sn18}) 
\begin{eqnarray}
ML = \frac{4\sqrt{2}}{\sqrt{1+\sqrt{1-2c}}}K(m).
\label{sn19}
\end{eqnarray} 

Taking the derivative with respect to parameter $c$ in (\ref{sn19}), it
is easy to show that $\frac{d(ML)}{dc} > 0$. In other words, $``ML"$ it is an 
increasing function of the parameter $c$, and so there exists one to one 
correspondence between $ML$ and $c$ (see Fig. $4$ below). Therefore, given an 
arbitrary $L$, there exists only one correspondent $c \in (0,\frac{1}{2})$ as 
is shown in Fig. $4$, and so there exists only one classical solution 
$\phi_{L}(x)$ satisfying Dirichlet boundary conditions.  

Taking the limit $c\rightarrow 0$ we have from (\ref{sn19}) 
\begin{eqnarray}
ML = 2\pi.
\label{sn20}
\end{eqnarray}
This relation shows that there exists a minimal value of $``ML"$ for the
solution (\ref{sn12}) satisfying Dirichlet boundary conditions. This is an 
interesting aspect of these solutions: once it is ``placed" in a box of size 
$L$, its mass must be greater than $\frac{2\pi}{L}$. On the other hand
if we fix $M$ as the mass of our field, it can not exist inside
a box with size smaller than $\frac{2\pi}{M}$. Of course if 
$L\rightarrow \infty$, the minimal mass $M\rightarrow 0$, and,
in this case, all positive mass values are allowed as it should
be in the DHN's model.

\newpage

\begin{figure}[ht]
\centerline{\psfig{figure=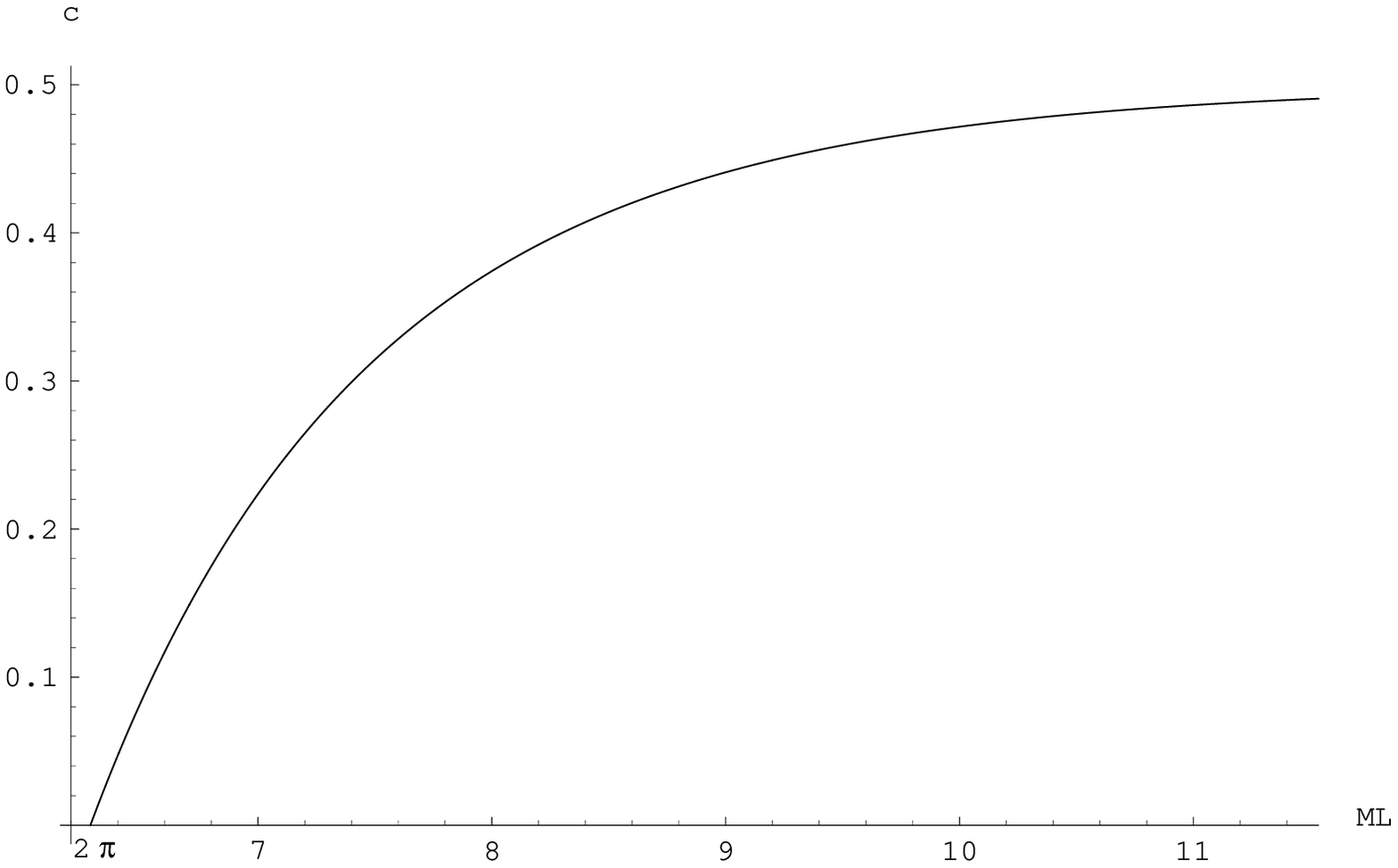,height=3.3in}}
%\caption{}
\end{figure}
\begin{center}
\noindent
\parbox{13cm}{Fig. 4. Constant of integration versus the product ML. 
The minimal value is reached at $ML = 2\pi$.}
\end{center}

\vspace*{0.5cm}

On the other hand, we can also define an expression that will be seen as 
``topological charge'' for the solution (\ref{sn12}). In the literature we find 
that the topological charge is defined as $^{6}$ :
\begin{eqnarray*}
Q=\frac{1}{2}\,\int_{-\infty}^{\infty}\,dx\,J^{0},
\end{eqnarray*}
where
\begin{eqnarray*}
J^{0}=\frac{\sqrt{\lambda}}{M}\varepsilon^{01}\,\frac{d\phi}{dx^{1}}, \ \ \ \ \ \ \ 
\mbox{and} \ \ \ \varepsilon^{ik} \ \ \mbox{is the antisymmetric tensor with} \ \
\varepsilon^{01}=1.
\end{eqnarray*}

In our case the ``topological charge'' will be given by
\begin{eqnarray*}
Q(L)=\frac{1}{2}\,\int_{-\frac{L}{2}}^{\frac{L}{2}}\,\frac{\sqrt{\lambda}}
{M}\, d\phi=\frac{1}{2}\,\frac{\sqrt{\lambda}}{M}\,\left(\phi(\frac{L}{2})-
\phi(-\frac{L}{2})\right).
\end{eqnarray*}

Using the solution (\ref{sn12}), it is easy to verify that
\begin{eqnarray}
Q(L)=\pm\,\frac{\sqrt{2c}}{\sqrt{1+\sqrt{1-2c}}}\,
{\rm sn}\left(\frac{ML}{2\sqrt{2}}\sqrt{1+\sqrt{
1-2c}},\frac{1}{-1+\frac{1+\sqrt{1-2c}}{c}}\right).
\label{sn21}
\end{eqnarray}

These are the ``topological charges'' for an arbitrary length $L$ of the
interval. It is seen from (\ref{sn17}) that with the Dirichlet boundary
conditions they are equal to zero.
For large lengths
($L=\infty$ or $c=\frac{1}{2}$) we recover the usual conserved charge
$Q$ of the Kink (AntiKink) solutions, that is
\begin{eqnarray*}
Q=\pm 1.
\end{eqnarray*}

In a similar way we now define
an expression for the ``total energy'' (the so called {\it classical mass} in 
the case of the Kink $^{7}$), associated to the solution (\ref{sn12}), 
that is
\begin{eqnarray}
{\cal M}(L) = \int_{-\frac{L}{2}}^{\frac{L}{2}} dx\,\epsilon_{c}(x),
\label{sn22}
\end{eqnarray}
where $\epsilon_{c}(x)$ is the ``energy density'' given by
\begin{eqnarray*}
\epsilon_{c}(x) = \frac{1}{2}(\partial_{x}\phi_{c})^2 + V(\phi_{c}) - 
V(\phi_{m}),
\end{eqnarray*}
and $\phi_{m}$ are the points of
minimum of the potential $V(\phi)$.

So, the ``energy density'' $\epsilon_{c}(x)$ is found to be 
\begin{eqnarray*}
\epsilon_{c}(x)=\frac{M^4}{\lambda}\left(\frac{c}{2}{\rm cn}^2
(\frac{Mx}{\sqrt{2}}\sqrt{1+\sqrt{1-2c}},\frac{1}{-1+\frac{1+\sqrt{1-2c}}{c}})
{\rm dn}^2(\frac{Mx}{\sqrt{2}}\sqrt{1+\sqrt{1-2c}},\frac{1}{-1+\frac{1+
\sqrt{1-2c}}{c}}) \right.
\end{eqnarray*}
\begin{eqnarray*}
- \ \frac{c}{(1+\sqrt{1-2c})}{\rm sn}^2(\frac{Mx}{\sqrt{2}}\sqrt{1+\sqrt{1-2c}},
\frac{1}{-1+\frac{1+\sqrt{1-2c}}{c}}) \ + \ \frac{c^2}{(1+\sqrt{1-2c})^2}\times
\end{eqnarray*}
\begin{eqnarray}
\left. {\rm sn}^4(\frac{Mx}{\sqrt{2}}\sqrt{1+\sqrt{1-2c}},\frac{1}{-1+\frac{1+
\sqrt{1-2c}}{c}}) \ + \ \frac{1}{4}\right).
\label{sn23}
\end{eqnarray}

Using the previous relation we obtain from (\ref{sn22})
\begin{eqnarray*}
{\cal M}(L)=\frac{M^3\sqrt{2}}{\lambda\sqrt{1+\sqrt{1-2c}}}\left\{
\frac{(1+\sqrt{1-2c})}{3}\left[ E\left( \mbox{am}(\frac{ML\sqrt{1+\sqrt{1-2c}}}
{2\sqrt{2}}),\frac{1}{-1+\frac{1+\sqrt{1-2c}}{c}}\right) \ - \right. \right.
\end{eqnarray*}
\begin{eqnarray*}
\left. \frac{1-2c+\sqrt{1-2c}}{1-c+\sqrt{1-2c}}\,
\frac{ML\sqrt{1+\sqrt{1-2c}}}{2\sqrt{2}}\right]
\ + \ \frac{2c}{3}(\frac{1}{2} \ + \ 
\frac{1+\sqrt{1-2c}-c}{(1+\sqrt{1-2c})^2})\times
\end{eqnarray*}
\begin{eqnarray*}
\left[ {\rm sn}(\frac{ML\sqrt{1+\sqrt{1-2c}}}{2\sqrt{2}},\frac{1}
{-1+\frac{1+\sqrt{1-2c}}{c}})\,
{\rm cn}(\frac{ML\sqrt{1+\sqrt{1-2c}}}{2\sqrt{2}},\frac{1}
{-1+\frac{1+\sqrt{1-2c}}{c}})\times \right.
\end{eqnarray*}
\begin{eqnarray*}
\left. {\rm dn}(\frac{ML\sqrt{1+\sqrt{1-2c}}}{2\sqrt{2}},\frac{1}
{-1+\frac{1+\sqrt{1-2c}}{c}})\right] \ - \ \frac{2}{3}\frac{(1+\sqrt{1-2c}-c)}
{(1+\sqrt{1-2c})}\left[ \frac{ML\sqrt{1+\sqrt{1-2c}}}{2\sqrt{2}} \ - \right.
\end{eqnarray*}
\begin{eqnarray*}
\left. E\left( \mbox{am}(\frac{ML\sqrt{1+\sqrt{1-2c}}}{2\sqrt{2}}),\frac{1}
{-1+\frac{1+\sqrt{1-2c}}{c}}\right) \right] \ + \ \left( \frac{(1-2c + 
\sqrt{1-2c})c}{3(-c+1+\sqrt{1-2c})} \ - \right.
\end{eqnarray*}
\begin{eqnarray}
\left. \left. \frac{2c(1+\sqrt{1-2c}-c)}{3(1+\sqrt{1-2c})^2} \ + \ 
\frac{1}{2}\right) \frac{ML\sqrt{1+\sqrt{1-2c}}}{2\sqrt{2}}\right\},
\label{sn24}
\end{eqnarray}
where $E(\gamma,\frac{1}{-1+\frac{1+\sqrt{1-2c}}{c}})$ is an elliptic integral
of the second kind, and 
$\gamma =\mbox{am}(\frac{ML\sqrt{1+\sqrt{1-2c}}}{2\sqrt{2}})$ is the so called 
amplitude of this integral $^{4}$.
Here the ``total energy '' (\ref{sn24}) depends on the geometrical parameter 
$L$. 

The ``energy density'' (\ref{sn23}) is shown in Fig. $5$ below for the several
values of $c$.

\begin{figure}[ht]
\centerline{\psfig{figure=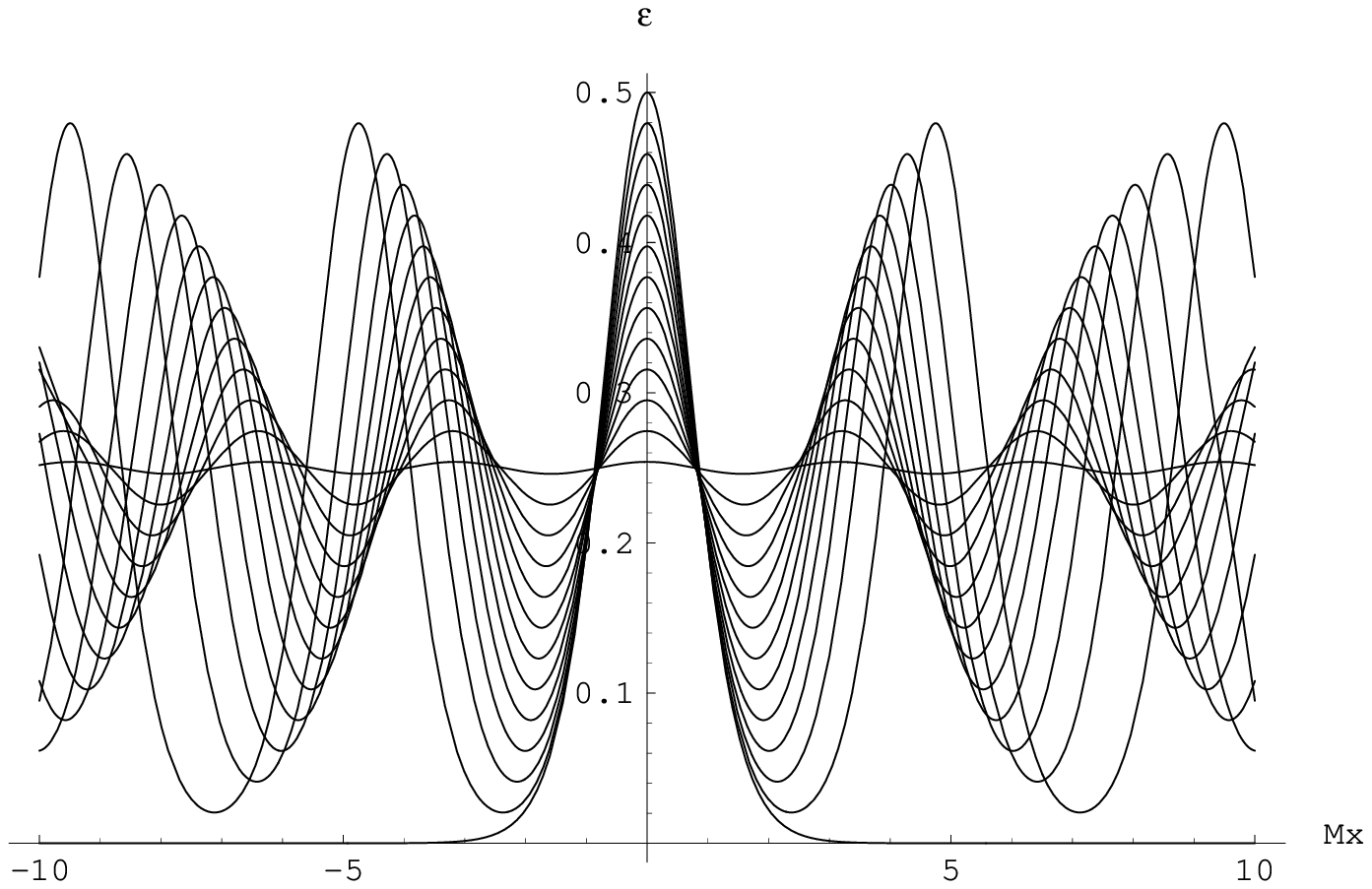,height=3.3in}}
%\caption{}
\end{figure}
\begin{center}
\noindent
\parbox{13cm}{Fig. 5. Energy density for a family of sn-type solutions (c = 0.008, 0.049, 
0.09, ..., 0.418, 0.459, 0.5). The curve with c = 0.5 corresponds to the energy 
density for DHN's Kink.}
\end{center}

\vspace*{0.5cm}

For $c=\frac{1}{2}$ in (\ref{sn23}) 
we recover the energy density of the Kink, i.e.
\begin{eqnarray}
\epsilon(x) = \frac{M^4}{2\lambda}\,sech^{4}(\frac{Mx}{\sqrt{2}}),
\label{sn25}
\end{eqnarray}
and from (\ref{sn24}) we recover the classical mass of the Kink 
\begin{eqnarray}
{\cal M} = \frac{2\sqrt{2}M^3}{3\lambda}.
\label{sn26}
\end{eqnarray}

\vspace*{0.7cm}

\noindent
{\Large \bf 3. Properties of {\rm cn}-Type Solutions in an
One-Dimensional Box}

\vspace*{0.7cm}

In this case for the different values of $c$ in (\ref{sn15}) there exist 
different values of the
``topological charge''. As in the last section we must discard the
solutions shown in Figs. 3.1 and 3.2  since they have also a  
discontinuity in the first derivative.

However, the  solutions shown in Figs. 3.3 and 3.4 are allowed since for a
given $L$ we can find such a $c$, that the solution exists in 
$(-\frac{L}{2},\frac{L}{2})$ and it has continuous derivative at all points
of the interval, including $x=0$. The charge is given by
\begin{eqnarray*}
Q(L)=\pm\sqrt[4]{2c}\,\frac{{\rm sn}
\left(\sqrt[4]{\frac{c}{2}}\frac{ML}{2},\frac{1}{2}(1+\frac{1}{\sqrt{2c}})\right)
{\rm dn}\left(\sqrt[4]{\frac{c}{2}}\frac{ML}{2},\frac{1}{2}(1+\frac{1}{\sqrt{2c}})
\right)}{{\rm cn}\left(\sqrt[4]{\frac{c}{2}}\frac{ML}{2},\frac{1}{2}(1+\frac{1}
{\sqrt{2c}})\right)}.
\end{eqnarray*}

When the condition (\ref{sn16}) of vacuum-vacuum transition in a finite domain 
is satisfied we obtain
\begin{eqnarray*}
Q=\pm 1.
\end{eqnarray*}

So, we recover the conserved charge $Q$ of the Kink (AntiKink) solutions
but now for a finite domain.

In the case of the ``energy density'' we have
\begin{eqnarray*}
\epsilon_{c}(x)=\frac{M^4}{\lambda}\left[\frac{c}{2}{\rm dn}^4
(\sqrt[4]{\frac{c}{2}}Mx,\frac{1}{2}(1+\frac{1}{\sqrt{2c}})) \ + \
\frac{c}{8}(1+\frac{1}{\sqrt{2c}})^2 {\rm sn}^{4}(\sqrt[4]{\frac{c}{2}}Mx,
\frac{1}{2}(1+\frac{1}{\sqrt{2c}})) \ + \right.
\end{eqnarray*}
\begin{eqnarray*}
c\,\frac{{\rm sn}^{4}(\sqrt[4]{\frac{c}{2}}Mx,\frac{1}{2}(1+\frac{1}{\sqrt{2c}})) 
\,{\rm dn}^{4}(\sqrt[4]{\frac{c}{2}}Mx,\frac{1}{2}(1+\frac{1}{\sqrt{2c}}))}
{{\rm cn}^{4}(\sqrt[4]{\frac{c}{2}}Mx,\frac{1}{2}(1+\frac{1}{\sqrt{2c}}))} \ - \  
\frac{c}{2}(1+\frac{1}{\sqrt{2c}})\,{\rm sn}^{2}(\sqrt[4]{\frac{c}{2}}Mx,
\frac{1}{2}(1+\frac{1}{\sqrt{2c}}))
\end{eqnarray*}
\begin{eqnarray*}
\times{\rm dn}^{2}(\sqrt[4]{\frac{c}{2}}Mx,\frac{1}{2}(1+\frac{1}{\sqrt{2c}})) \ + \
c\,\frac{{\rm sn}^{2}(\sqrt[4]{\frac{c}{2}}Mx,\frac{1}{2}(1+\frac{1}{\sqrt{2c}}))
\,{\rm dn}^{4}(\sqrt[4]{\frac{c}{2}}Mx,\frac{1}{2}(1+\frac{1}{\sqrt{2c}}))}
{{\rm cn}^{2}(\sqrt[4]{\frac{c}{2}}Mx,\frac{1}{2}(1+\frac{1}{\sqrt{2c}}))} \ - \
\end{eqnarray*}
\begin{eqnarray*}
\frac{c}{2}(1+\frac{1}{\sqrt{2c}})\,\frac{{\rm sn}^{4}(\sqrt[4]{\frac{c}{2}}Mx,
\frac{1}{2}(1+\frac{1}{\sqrt{2c}}))
\,{\rm dn}^{2}(\sqrt[4]{\frac{c}{2}}Mx,\frac{1}{2}(1+\frac{1}{\sqrt{2c}}))}
{{\rm cn}^{2}(\sqrt[4]{\frac{c}{2}}Mx,\frac{1}{2}(1+\frac{1}{\sqrt{2c}}))} \ - \
\frac{\sqrt{2c}}{2}\times
\end{eqnarray*}
\begin{eqnarray*}
\left. \frac{{\rm sn}^{2}(\sqrt[4]{\frac{c}{2}}Mx,\frac{1}{2}(1+\frac{1}
{\sqrt{2c}}))\,{\rm dn}^{2}(\sqrt[4]{\frac{c}{2}}Mx,\frac{1}{2}(1+\frac{1}
{\sqrt{2c}}))}{{\rm cn}^{2}(\sqrt[4]{\frac{c}{2}}Mx,\frac{1}{2}(1+\frac{1}
{\sqrt{2c}}))}\ + \ \frac{1}{4} \right].
\end{eqnarray*}

For $c=\frac{1}{2}$ we recover once more the energy density of the Kink 
(\ref{sn25}).

On the other hand it is not difficult to obtain an expression for the
``total energy'' associated to these solutions using the previous relation in 
(\ref{sn22}). Again for $c=\frac{1}{2}$ we recover the classical mass
of the Kink, i.e., (\ref{sn26}).

A similar calculation can be made considering periodic or
anti-periodic boundary conditions for the solution (\ref{sn12}).
Imposing periodic boundary conditions at $x=\pm\frac{L}{2}$, namely
\begin{eqnarray*}
\phi_{c}(-\frac{L}{2})=\phi_{c}(\frac{L}{2}),
\end{eqnarray*}
we obtain (both for the positive and negative signs in
(\ref{sn12})), the same relation (\ref{sn17}) and, therefore, all 
results of the Dirichlet case are valid.

The anti-periodic boundary conditions 
$\phi_{c}(-\frac{L}{2})=-\phi_{c}(\frac{L}{2})$ \ 
are satisfied automatically
for $\forall c\in (0,\frac{1}{2}]$ and $\forall
L\in (0,\infty)$ since the solutions (\ref{sn12}) are odd functions 
(we took $x_0 =0$). So, this boundary condition does not furnish any
constraint like those which we have for a Kink.

\vspace{0.7cm} 

\noindent
{\Large \bf 4. Conclusion}

\vspace*{0.7cm}
 
In this work we found the general solution of the scalar wave
equation for the potential
$V(\phi) = -\frac{1}{2}M^2\phi^2 + \frac{\lambda}{4}\phi^4$ in a box in a two 
dimensional space-time. These solutions are the well known Jacobi Elliptic 
Functions. We showed that solutions given by Eqs. (\ref{sn6}), (\ref{sn9}) and 
(\ref{sn12}) can 
not satisfy the vacuum-vacuum boundary conditions in a finite domain, but 
those solution given by Eq. (\ref{sn15}) can satisfy
 these conditions. 
Finally, we showed that there exists one classical real solution
$\phi_{L}(x)$ in an interval with size $L$ satisfying Dirichlet's boundary
conditions. In this case the product $``ML"$ has a minimal value $2\pi$. 
This implies that a solution corresponding to a mass $M$ can not exist in a 
cavity with length smaller than $\frac{2\pi}{M}$.
As in the Casimir Effect these solutions show that physical properties of some
systems can change drastically when placed in cavities $^{2}$.

%A interesting question
%is about the existence of Kink solutions on boundaryless manifold such as
%spheres, torus, etc. An possible aplication for these solutions would be in
%cosmology where ones could consider Kinks on $S^{3}$ as seeds for galaxy
%formation.

We have defined expressions for the ``topological charge'' 
(\ref{sn21}), ``total energy'' (\ref{sn24}) and ``energy density'' (\ref{sn23}) 
associated to the elliptic {\rm sn}-type solutions in a finite domain
and shown that for large lengths $(L=\infty \ \ \mbox{or} \ \ 
c=\frac{1}{2})$ we recover the conserved charge, classical mass and energy 
density of the Kink. Also, for periodic boundary 
conditions all results are valid obtained with Dirichlet boundary conditions. 
For anti-periodic boundary conditions it is easy to see that all solutions of 
the class defined by Eq. (\ref{sn12}) are allowed. These solutions, however, 
do not satisfy the asymptotic vacuum-vacuum boundary conditions, 
unless $c = \frac{1}{2}$ and $L=\infty $.

\vspace{1cm} 

\noindent
{\large \bf
Acknowledgments}
\vspace*{0.2cm}

\noindent This work was supported in part by  FAPESP (Funda\c{c}\~ao de Amparo 
\`a Pesquisa do Estado de S\~ao Paulo), Brazil. 

%\newpage
\vspace*{1cm}

\noindent
{\large \bf
References}
\vspace*{0.2cm}
\begin{flushleft}
\noindent 
1. {R. Dashen, B. Hasslacher and A. Neveu, {\em Phys. Rev. 
{\bf D10}, 4131 (1974)}}.\newline
2. {V. M. Mostepanenko and N. N. Trunov, {\em The Casimir Effect
and its Applications} (Oxford
\hspace*{0.35cm} Univ. Press, Oxford, 1997)}.\newline
3.  {I.S. Gradshteyn and I.H. Ryzhik. {\em Table of Integrals,
Series and Products}, Academic Press, \hspace*{0.35cm} New York (1980)}.\newline
4. {M. Abramowitz and I. A. Stegun. {\em Handbook of Mathematical
Functions} (Dover \hspace*{0.35cm} Publications, 
INC., New York, 1972)}.\newline
5.  {P. G. Drazin and R. S. Johnson, {\em Solitons: An Introduction}
(Cambridge Univ. Press, \hspace*{0.35cm} Cambridge, 1989).}\newline
6.  {Lewis H. Ryder, {\em Quantum Field Theory}
(Cambridge Univ. Press, Cambridge, 1988).}\newline
7.  {R. Rajaraman, {\em Solitons and Instantons: An Introduction to Solitons
and Instantons in \hspace*{0.35cm} Quantum Field Theory} (North Holland, 1982).}

\end{flushleft}

\end{document}